\documentstyle[12pt]{article}
\begin{document}
\begin{picture}(10,30)(0,0)
\put(220,15) {SNUTP 94-116}
\put(220,0) {INP MSU - 94 - 36/358}
\end{picture}
\begin{center}
{\large \bf CompHEP  \\
Specialized package for automatic calculations \\ of elementary particle decays
and collisions \footnote{ The talk given at the Korean Physical Society meeting
on October 21, 1994}}
\end{center}

\begin{center}
{\bf E.E. Boos, M.N.Dubinin, V.A. Ilyin, A.E. Pukhov, V.I. Savrin}\\
Institute of Nuclear Physics, Moscow State University \\
119899 Moscow, Russia
\end{center}
\begin{abstract}
  At present time when a new
generation of TeV colliders are beginning to operate one needs to calculate
cross-sections for a great number of various reactions. Such calculations
are united in the framework of the collider physical program, providing
definite predictions how to detect the signatures of the new physics and
separate them from the background.
         The CompHEP package was created for calculation of decay and
     high energy collision processes of elementary particles in the
     lowest order (tree) approximation.
The main
 idea
put into the CompHEP  was to make available passing from the lagrangian to the
 final
distributions effectively with the high level of automatization what is
extremely
needed in collider physics.
\end{abstract}

\begin{center}
{\bf  1. Introduction}
\end{center}

 CompHEP project was started in
     1989 by group of physicists and programmers  from the
Institute of
 Nuclear Physics, Moscow
State University.
    Twelve researchers (P. Baikov, E. Boos, M. Dubinin,
   H. Eck, V. Edneral, D. Kovalenko,
   A. Kryukov, A. Pukhov, A. Semenov, S. Shichanin, P. Silaev,
A. Taranov)
   took part in the development of the  project coordinated by
   Dr. V.Ilyin and Prof. V.I.Savrin.  The main steps of the CompHEP development
are published
   in Ref.\cite{r1}. The project was supported by
   Russian
   State Program on High Energy Physics, RFFI (93-02-14428), ISF (M9B000),
   INTAS (93-1180), Japan Society for the Promotion of Science, Japanese
   companies KASUMI Co, Ltd and SECOM Co, Ltd. and Royal Society (UK).

The first versions of the CompHEP package were written in Turbo Pascal for
IBM
 compatible PC.
In 1992 this package was rewritten in the C programming
language and now the installation on UNIX workstations is available.
At present time there are some  versions for  different platforms: HP Apollo
9000, IBM RS 6000, DECstation 3000, Sparc station, Silicon Graphics.
Adaptations for these platforms were  done during the visits of the CompHEP
group members in  KEK (Japan), Seoul National University (Korea),
    University "La Sapienza" (Rome, Italy), University of Sao Paulo (Brazil),
     DESY (Germany).

      CompHEP is a menu-driven system
     with the mixed text/graphical
     output of information and the context HELP facility.
     The notations used in CompHEP
     are very similar to those used in particle physics.
     The present version has 4 built-in physical models. Two of them are
versions of the Standard
  Model (SU(3)xSU(2)xU(1)) in the unitary and 'tHooft-Feynman gauges. The user
can change
 interaction vertices and model
 parameters. A creation of a new particle interaction model by the user is
 possible.
        In the present version   polarizations are not taken into account.
     Averaging over
   initial and summing over final polarizations are performed
   automatically.

   The general structure of the  CompHEP package is represented  in Fig.1.
It
consists of two parts: the symbolical part and
  the   numerical one. The main tasks solved by
     the symbolical part are :

          1.  to select process  by specifying  {\it in-}  and {\it out-}
particles
          for decays of $1 \to 2, \ldots ,1 \to 5$ type and collisions of
          $2 \to 2, \ldots, 2 \to 4$ type;

          2.  to generate and display Feynman diagrams;

          3.  to delete some number of diagrams from the further
consideration;

          4.  to generate and display squared Feynman diagrams
             (corresponding to squared S-matrix elements);

          5. to calculate analytical expressions corresponding to squared
             diagrams with the help of the fast built-in symbolic calculator;

          6. to perform numerical calculations  for $1 \to 2$ and $ 2 \to 2$
       processes, to show plots of angular distributions and cross sections;

          7.  to save symbolic results corresponding to the squared diagrams
             calculated in the REDUCE and MATHEMATICA codes for further
             symbolical manipulations;

          8. to generate the optimized FORTRAN code for the squared matrix
             elements for  further numerical calculations;

     The numerical part of the CompHEP package is written in  FORTRAN. It
uses
     the CompHEP FORTRAN output, the  BASES\&SPRING package \cite{r2} for
     Monte-Carlo integration and event generation. The main tasks solved by
     the numerical  part are :

    1. to choose phase space kinematical variables;

    2. to introduce kinematical cuts over any squared momentum transfers
and
    squared masses for any groups of outgoing particles;

    3. to perform  a regularization to  remove  sharp peaks in the squared
matrix
     elements;

    4. to change the BASES parameters for  Monte-Carlo integration ;

    5. to change numerical values of  model parameters;

    6.  to calculate distributions, cross-sections or
    particle widths by the Monte-Carlo method;

    7. to perform the same intergration taking into account structure function
        for incoming particles;

    8. to generate events and to  get histograms simulating the signal
       in real experiment.

\begin{center}
\bf  2. Menu system of the CompHEP  symbolic part
\end{center}

The  CompHEP is a menu-driven system.
The user  can select a menu position
     with the help of the ARROW keys.
     The input of the selected position is performed by pressing the ENTER key.
     Before pressing ENTER one can press F1 key (HELP) to get information
about the
     selected menu position.
     To  return to the previous level menu  one should  press the ESC or
BACKSPACE keys.

 The menu titles of the CompHEP symbolic
part are shown in Fig.2.

{ \bf Menu 1  } (models)

This menu gives the user a possibility to select a model  of
elementary particle interaction.

1. {\it  QED }    denotes  Quantum Electrodynamics.

2. {\it  Fermi model}   denotes  QED with the four-fermion weak interaction
model.
         The interaction of fermion currents is implemented through
          auxiliary particles (W and Z bosons) having constant
          propagators.

3. {\it  St.model (unit. gauge) }   denotes the Standard Model with the
electroweak
     sector  in the unitary (physical) gauge and QCD in the Feynman
gauge.

4. {\it  St.model (Feyn. gauge)}  denotes the Standard Model with the
electroweak
     sector in the 'tHooft - Feynman gauge and QCD in the Feynman gauge;

 {\it N E W \quad M O D E L} is an option for creating a new
       physical model. The new model is created as a copy
       of some old model with some new name. The {\it Edit model}
       option of Menu 3 can be used to insert changes.
         If the user chooses  this menu line  he will be asked
about a new model name and about a template for model. To choose the
template the list  of the existing models appears.

       After a model selection the user  gets to  Menu 2 for further
processing.

{\bf Menu 2 }

1. {\it Enter process } is an option for entering the process from the
keyboard (see Section 4).  If the  input is correct, CompHEP constructs the
corresponding
 Feynman diagrams and the user enters  Menu 4.

2. {\it Edit model }  option allows the user to view and change the current
physical model. The user enters  Menu 3 to insert changes into model
tables.

3. {\it Delete changes } option removes the model
created with the help of the {\it N E W \quad  M O D E L } option from
 Menu 1 or restores  the built-in model if it has been changed
with the help of the {\it Edit model } option.

{\bf Menu 3 } (edition of models)

 A physical model in CompHEP consists of four tables:

1.{\it Variables }  (the list of independent parameters)

2.{\it    Constraints } (the list of dependent  parameters)

3.{\it    Particles }   (the list of particles)

4.{\it    Lagrangian }  (the list of vertices)

By this Menu the user selects a table for  changes (see Section 3). After
changes the user should
 press  the ESC  key to  leave this Menu. If the tables are changed the user
  will be asked:

 {\it   Save changes Y/N ? }

If the user's answer is "N" then the changes will be forgotten. In other
case
 CompHEP  checks the  new version of the model. If the version is correct
it is saved into the user's directory  {\it Models }.  Otherwise the
message about an error appears on the screen.

{\bf Menu 4 }

1. {\it Squaring }  generates  diagrams for squared S-matrix elements.

2. {\it View diagrams}  displays  graphic  representation  of the constructed
                  Feynman diagrams (see Fig. 3).
                  Here the user has a possibility to exclude some diagrams from
                  further processing.

When squaring is finished  Menu 5  is displayed.

{\bf Menu 5}

1. {\it View squared diagrams } shows  graphic  representation  of the
constructed squared  diagrams. Here the user again has a possibility to
exclude some diagrams from further processing.

2.      {\it Symbolic calculations }   starts symbolic
                  calculations of  the squared diagrams generated. In
                  CompHEP the calculation is done by means of the
built-in   symbolic manipulation package.

3.  {\it  Write results } - gives the possibility to write obtained
analytical
                  result for squared diagrams on the hard disk in different
                  formats, the corresponding files are placed in the
                  subdirectory RESULTS. This option is available after
               the   symbolic calculation only. Menu 6  for different
                  output formats is displayed.

4.  {\it REDUCE program }  generates for each squared diagram
                   a source code in the REDUCE format for the following
                   calculation of the matrix element squared by means of
                   REDUCE.  This option can be used to check the build-in
		  symbolic calculator.

5.  {\it  Numerical calculator} fulfills numerical calculations for the
simplest $1 \to 2$ and $2 \to 2 $
processes. Calculated numerical values of widths, distributions or cross
sections
are displayed on the screen, also angle distributions. In the case of
infrared singularities the message
{\it Division by zero } is written. Menu 7 appears for further processing.

6.  {\it  Enter new process}  returns to the
                   Menu 2 for entering a new collision or decay
                   process.

7.  {\it   Interface} gives a possibility to incorporate the CompHEP session
with
            the work of other external packages. The Menu of external
packages
            appears. The first position of this Menu is used to start the
numerical
            part of the CompHEP package (see Section 5). Other positions are
free.

{\bf Menu 6}

 1.  {\it FORTRAN code}   writes down symbolical results in the directory
RESULTS
 as  a FOFTRAN code for further numerical calculations.

  2/3. {\it REDUCE code / MATHEMATICA code }   writes down symbolical results
  in the directory RESULTS as  a REDUCE/MATHEMATICA code for further
  symbolical manipulations.

{\bf Menu 7}

1. {\it  View/change data} shows the table with parameter names
and numerical values
 on the screen.  It gives the  user a possibility to change a numerical value
for any parameter. After changes one should press ESC to leave the parameter
table. The cross section (width) will be recalculated  automatically.

2. {\it Set angular range}  (for $2\to 2$ processes only) allows
 to  set {\it min } and {\it max} values for cosine of scattering
angle in the center of mass  reference system.

3. {\it Set precision }  sets a precision for the numerical calculation.

4. {\it Angular dependence} allows to draw (also save into the file)
the differential cross section as a function of cosine of scattering angle
in the center of mass frame.  Menu 8 is called  for further work.

5. {\it  Parameter dependence } allows one to get distributions
for any parameter involved into the process under consideration.
For the $1\to 2$ process the parameter dependence of the width can be
displayed.
For  the $2\to 2$ process the user can construct plots both  for the
cross section and
 for the asymmetry (the forward - backward cross section difference). This
selection is realized  by Menu 9, the user getting a table
  to select parameters. Then  the user is
asked about minimal and maximal values of these parameters and
about scales (normal / logarithmical) for parameters.
After finishing the  calculations  the user gets Menu 10.

{\bf Menu 8}

 1. {\it Show plot} displays  on the screen the calculated  angular
dependence.
 Two ordinates  scales are available: normal and logarithmic ones.

 2. {\it Save results in a file}  writes  down the calculated angular
dependence in a file. The file is
 created in the RESULTS directory. The message about the file name
 appears on the screen.

 3. {\it Recalculate} recalculates the differential cross section for a
given  number of points and with higher precision.

{\bf Menu 9 }: (see the item 5 in Menu 5)

{\bf Menu 10}

 1. {\it Show plot} displays  on the screen graphical plots for the
 total cross section (asymmetry)
 as a function of selected parameter.  For positive functions
 two ordinate  scales are available: normal and logarithmic ones.

 2. {\it Save results in a file} writes down the calculated  parameter
dependence
 in a file. The file is
 created in the RESULTS directory. The message about the file name
 appears on the screen.

\begin{center}
{\bf 3. Physical model structure in  CompHEP.}
\end{center}

The  information about a physical model in CompHEP is stored
in four tables. They are

{\it    Parameters}  - the table of independent parameters,

{\it    Constraints } -  the table of dependent  parameters,

{\it    Particles }   - the table of particles,

{\it    Lagrangian }  - the table of vertices.

         All independent parameters used in  calculations  are  described in
the    {\it  Parameters} table:

\begin{center}
\begin{tabular}{|l|l|l|}
\hline
 Name   & Value   & Comment      \\
\hline

 EE     & 0.313   & electromagnetic coupling constant \\
 GG     & 1.373   & strong coupling constant          \\
 SW     &0.48     & sine of the Weinberg angle         \\
 MZ     &91.16    & Z-boson  mass                     \\
 wZ     &2.53     & Z-boson  width                    \\
 wW     &2.25     & W-boson  width                    \\
 ...    & ...     &  ... \\
 \hline
 \end{tabular}
 \end{center}

       {\it  Name }  means the name of the parameter used for a further model
			definition  and a symbolic result representation.
      {\it   Value} is a numerical value of the parameter used for numerical
                   calculations.
       {\it  Comment} is a  commentary what the corresponding parameter means.

   The table {\it Constraints}
 describes  variables that are  functions of parameters:

\begin{center}
\begin{tabular}{|l|l|l|}
\hline
  Name & Expression & Comment             \\
\hline
 CW    &Sqrt(1-SW**2)  & cosine of the Weinberg angle\\
  MW   & MZ*CW         & W boson mass         \\
 ... &  ... & ...  \\
 \hline
 \end{tabular}
 \end{center}

Here {\it Name }  means the name of a new  variable, {\it Expression }
defines  a function  of independent parameters and dependent variables
defined above.

         The {\it Particles} table consists of 8 fields:

\begin{center}
\begin{tabular}{|l|l|l|l|l|l|l|l|}
\hline
 Full name & P  & aP & 2*Spin & Mass & Width & Color & Aux \\
\hline
gluon        &G  &G  &2     &0     &0     &8    &G   \\
electron     &e1 &E1 &1     &0     &0     &1    &    \\
e-neutrino   &n1 &N1 &1     &0     &0     &1    &L   \\
u-quark      &u  &U  &1     &0     &0     &3    &    \\
W-boson      &W+ &W- &2     &MW    &wW    &1    &G   \\
Z-boson      &Z  &Z  &2     &MZ    &wZ    &1    &G   \\
 ...       & ...& ...& ...    & ...  & ...   &  ...  & ... \\
 \hline
 \end{tabular}
 \end{center}

  {\it Full name} denotes the  name of the particle. This field is a
comment only.

  {\it    P } and {\it aP}  contain designations of the
                   particle and antiparticle correspondingly. If the
                   antiparticle is identical to the particle  {\it aP}
must    be equal to {\it P}.

   {\it  2*Spin}  denotes the doubled spin of the particle.  Only
		     0, 1 and 2 values  are
available here (scalar, spinor and vector cases).

{\it    Mass}  and {\it Width} are  particle mass and width identifiers.
		       The values of these symbolic
		         parameters have to be entered in the tables
{\it Parameters} or
				 {\it Constrains}. It is possible to enter zero in these fields.

{\it    Color}  denotes  a dimension of the  color SU(3) group
representation:
                   3 for quarks, 8 for gluons, 1 for colorless particles.

{\it     Aux}  is  a  special  auxiliary  field:
               {\it L/R} are marks for left/right massless spinor particles,
           {\it G} is a mark for a vector particle to be treated as a
		   gauge field  in the 'tHooft - Feynman gauge,
          asterisk "*" is a mark for a massive  particle that will have
the constant
   propagator $\frac{1}{M^2}$. This option is used for implementation of the
		   four-fermion vertex in the Fermi model of electroweak interaction.

         Propagators of any particles are generated by CompHEP
     automatically.  CompHEP uses the standard form for scalar,
     spinor and vector particle propagators.
	 However the field "Aux" provides the user with some possibility to change
	 propagators.

 The table {\it Lagrangian} includes
 the list of vertices.
{\small
\begin{center}
\begin{tabular}{|l|l|l|l|l|l|}
\hline
   A1 & A2 & A3 & A4 & Factor         & Lorentz part            \\
\hline
W+ &W- &Z  &   &EE*CW/SW          &m1.m2*(p1-p2).m3+m2.m3*(p2-p3).m1 \\
   &   &   &   &                  &+m3.m1*(p3-p1).m2 \\
N1 &e1 &W+ &   &EE/(2*Sqrt2*SW)   &G(m3)*(1-G5)                           \\
E1 &n1 &W- &   &EE/(2*Sqrt2*SW)   &G(m3)*(1-G5)                           \\
U  &u  &G  &   &GG                &G(m3)                                  \\
 ...&...&... &... &...                     &...                    \\
\hline
\end{tabular}
\end{center}
}
      The first four fields ( A1, A2, A3, A4 ) contain the names of
     interacting particles.  The last two fields {\it Factor}  and  {\it
Lorentz part} define
     the vertex itself.
{\it Factor} is a scalar factor dependent only on  model variables.
It can not include summation of terms but only a single term.

{\it Lorentz part} depends on model variables, Lorentz momenta (p1, p2, p3 and
p4)
and Lorentz indices (m1, m2, m3 and m4) corresponding to the particles
listed in the first four fields: p1 and m1 for the first particle etc.

       Here dot symbol "." is used for a  scalar product of
Lorentz momenta ($p1.p2 = g_{\mu\nu} p1^{\mu} p2^{\nu}$) or indices
($p1.m1 = p1_{\mu_1}, \; m1.m2 = g_{\mu_1\mu_2}$);
G denotes the Dirac's gamma matrix ($G(m3)=\gamma^{\mu_3}$);
G5 is the $\gamma^5$ matrix ($\gamma^5 \gamma^5 = 1$)

The normalization condition is the following. Let S is the action. Then
for a three particle vertex

$$ \frac{\delta^3 S}{\delta A1_{[m1]}(p1) \delta A2_{[m2]}(p2)
\delta A3_{[m3]}(p3)} =(2 * \pi)^{-2} * \delta (p1+p2+p3) $$
$$ * [\gamma^0] *  ColorStructure * Factor * LorentzPart $$

     Here {\it ColorStructure } is a QCD  color structure. It is
     generated automatically  from a particle color weight.
     It is  impossible to
     introduce into the table the 4-gluon vertex explicitly in this way.
	  This vertex (and corresponding diagrams)
	 is generated automatically in some special way during the symbolic
	 calculation of squared diagrams.

Optional  $\gamma^0$ appears for a vertex with fermions to get a  Lorentz
covariant object. The Majorana representation for $ \gamma$-matrix is
implied.

\begin{center}
{\bf     4. Entering a  process in CompHEP.}
\end{center}

         After the input of the command {\it Enter process} of the Menu 2
          the list of particles will
     be shown contained in the model together with the corresponding
     notation conventions. The notation of  an antiparticle is shown in
     parenthesis after the notation of  a particle. For example  in the
    case of  the Standard Model:

\begin{center}
\begin{tabular}{|l|l|l|}
\hline
A(A) -photon &  G(G) -gluon & e1(E1) -electron \\
n1(N1) -e-neutrino  & e2(E2) -muon & n2(N2) -mu-neutrino \\
e3(E3) -tau   &  n3(N3) -tau-neutrino & u(U) u-quark \\
d(D) -d-quark &  c(C) c-quark & s(S) s-quark \\
t(T) t-quark  &  b(B) b-quark & H(H) -Higgs \\
W+(W-) W-boson & Z(Z) Z-boson   &  \\
\hline
 \end{tabular}
 \end{center}

In the down part of the screen the prompt {\bf Enter process } appears.
      The     syntax for the input is the following:
              first {\it in}-particles separated by comma, then
              arrow "$->$", and then
              {\it out}-particles separated by commas.
          For example,

          {\bf Enter process: $u,U->G,G$ } \\
     denotes the process of annihilation of u-quark and u-antiquark
     into two gluons.

         One can also construct inclusive processes. For example,

{\bf Enter process: $ u,U->G,G,2*x $ }\\
     is the request to construct all processes of
        annihilation of u-quark and u-antiquark
     into two gluons  and two arbitrary particles.

         If the program finds the unknown symbol  among the
     in-particles this symbol is considered as a composite particle. For
     instance, after the input

     {\bf Enter process: $ e,p->3*x $ }\\
      the question appears:

     {\bf  Is 'p' a composite  particle Y/N  ? } \\
     If one answers 'Y' he  will be prompted to specify the proton
     structure (list of partons).  A possible input is

{\bf      'p' consists of: $ u,U,d,D,G $ }

         If one  enters the collision process (not particle decay) then
     he will be asked about total energy of colliding particles
     in the  center-of-mass system (in GeV):

     {\bf Enter Sqrt(S) in GeV: {\it 300 } }

         On the next step the user  will be asked if it is necessary to
     generate diagrams with all possible intermediate (virtual)
     particles. These restrictions are used to exclude diagrams which
     are suppressed due to a large particle mass  or a small coupling
     constant or due to some other reasons. For example:

    {\bf   Restrict  virtual  particles: $  W+<2 $ }

         This particular restriction means that only diagrams with no
     more than one virtual W-boson will be generated.

     Several restrictions separated with commas are allowed.

\begin{center}
{\bf 5. Numerical part of the CompHEP package}
\end{center}

This program provides the user a possibility to prepare the
CompHEP FORTRAN output for a further numerical integration over
phase space and then to carry out this integration in a user-friendly
manner. This program provides also an interface with the
Monte Carlo integration package BASES and the
event generator SPRING.

  The user has a possibility to calculate
widths for decays or cross sections  for collisions.
Then one can  fill  histograms for different distributions.
This program  does not allow to make a
     summation over types of {\it in-} or {\it out-} particles.

     There are two modes of the program run available: interactive
     and batch ones. In the interactive mode the program is a menu-driven
system.
The menu titles of the CompHEP numerical part is reproduced in Fig.4.
To select the menu position the user should  type its number and press
the ENTER key.  To get HELP about any menu position the user should enter
h\#, where \# is the position number, or h for general help for this menu.

    The {\it\bf Calculation} position of {\bf Main menu} starts
a calculation of the cross-section (for collisions)
or the width (for decays) by BASES.
The program  operation is organized as a sequence of  work
sessions. The session number is displayed and automatically increasing after
each
sequential  Monte Carlo calculation.

After completing  the BASES calculation the {\it\bf Event generator}
submenu (the
interface with the SPRING program) will appear
if the event generator has been  switched on in the
submenu {\it\bf MC parameters}:

\begin{center}
\begin{tabular}{|ll|}
\hline
 \multicolumn{2}{|c|}{Event generator menu } \\
  \hline
    1: Start generator  &  2: Number of events = $10000$  \\
    3: View current `hst` file  &                  \\
\hline
\end{tabular}
\end{center}

The file {\sf hst.\#} contains the report of the SPRING run with histograms
initialized for  filling by the user.

 Other {\bf Main menu}  positions call submenus for setting conditions of the
 Monte Carlo integration.
Below we give a brief description of the submenu titles.

The {\bf IN state} submenu  serves for preparing the initial
state of collision processes.
  It is possible:
  \begin{itemize}
  \item to change $\sqrt{s}$ -- summary energy of {\it in}-particles in CMS,
  \item to switch on a structure function option.
  \end{itemize}

  The {\it\bf Model parameters} submenu
     allows one to change any physical model parameters
involved in the process. New values will be saved in
 a file.

The {\it\bf Invariant cuts} submenu
   allows to introduce kinematical cuts (lower and upper
   bounds) for  arbitrary squared
combinations of Lorentz momenta of {\it in-} and {\it out-} particles.
These cuts are written down in the table. For example:

\vspace{.5cm}
\vbox{
\begin{tabular}{|c|c|c|c|c|}
\hline
         N & MIN VALUE &  INVARIANT           & MAX VALUE & STATUS \\
           &  [GeV**2] &                      &  [GeV**2] &        \\
        \hline
         1 &           &  (-p1+p3)**2         & $<$ -1.000 & HARD   \\
         2 &           &  (-p2+p4)**2         & $<$ -1.000 & HARD  \\
        \hline
\end{tabular}
}

\vspace{.5cm}

  The {\it\bf Kinematics} submenu is used for defining
integration variables. First one has to choose the
necessary number of independent kinematical variables,
express
all scalar products of particle momenta and other Lorentz vectors
(if they are) via these integration variables, evaluate squared matrix element
 for the current phase space point
and multiply squared matrix element by the corresponding Jacobian.
This menu allows the user to define kinematics in the most convenient way for
the further integration over phase space.

The  scheme of kinematical variable selection are fixed in the table.
 For example:

\vspace{.5cm}
\vbox{
\begin{tabular}{|c|c|c|c|c|}
\hline
Decay &    In    & Out 1  & Out 2 & Pvect \\
\hline
 1  &   p1+p2  &   p3  &  p4+p5   &  -p1  \\
 2  &   p4+p5  &  p4   &   p5     &   -p2 \\
\hline
\end{tabular}
}
\vspace{.5cm}
  The {\it\bf MC parameters} submenu allows
the user to change some BASES \cite{r2} parameters which are engaged  in this
program.   There are  two loops of the BASES calculation.
 The  first one consists of iterations
  with  adaptation of the grid  from iteration to iteration.
The second loop includes iterations with the fixed grid to accumulate necessary
statistics. So,

\begin{description}
\item[Ncall] -- number of Monte-Carlo sample points for one iteration;
\item[Itmx1] -- maximal number of iterations with the grid
adaptation (1st loop);
\item[Acc1] --  limit for the calculation accuracy in \% (1st loop);
\item[Itmx2] -- maximal number of iterations with the fixed grid (2nd loop);
\item[Acc2] --  limit for the calculation accuracy in \% (2nd loop).
\end{description}

Also it is possible to  switch on the {\bf Event generator} SPRING and
define a number of events to be generated.

 The {\it\bf Regularization} submenu is used to
transform integration variables
for representing the  integrand as a smooth function.

In some cases the matrix element squared has very strong singularities and
 the  Monte Carlo integrators are not efficient enough.
  The typical example of the singular integration is given by the process

\centerline{\sf e1,A $\rightarrow$ e1,Z,H,}
\noindent
with the singularity appearing from the exchange of t-channel electron.

  For a reliable evaluation of such singular integrals this program has the
  special
option which can be activated in the menu.  Certainly this option is
available only if there is the correspondence between singularities and a
set of the integration variables. So, to make the regularization it could be
necessary to change kinematics.

The introduction of kinematical
regularizations speeds up the convergency of the Monte Carlo integration.
The invariants over which the regularization is made are written down
in the table. For example:

\vspace{.5cm}
\vbox{
\begin{tabular}{|c|c|c|c|c|}
\hline
 N &  INVARIANT  &  MASS [GeV]  &  WIDTH [GeV]   & STATUS \\
\hline
    1 & (-p1+p3)**2         &  0             &             &  ON    \\
    2 & (-p2+p4)**2         &  0             &    &  ON    \\
\hline
\end{tabular}
}
\vspace{.5cm}
 The {\it\bf Task formation} submenu provides the
following options :

\begin{itemize}
\item  to collect results of the calculation in the table(s) with any physical
   parameters as a table argument;
\item  to prepare the task for batch mode calculation;
\item  to set default session parameters.
\end{itemize}

 The {\it\bf View results} submenu allows the user
 to view any output files containing results of
   cross section (or width) calculation, a report on the process of MC
   integration, histograms.

   As a result of the calculation for each work session the program creates
three  output files:

\centerline{\sf res.\#, \hspace{0.2cm} prt.\# \hspace{0.2cm}
  and \hspace{0.2cm} hst.\#.}

Here \# denotes the number of session.
The file {\sf res.\#} contains a result of the calculation with a list of model
  parameters used.

The file {\sf prt.\#} is a copy of the screen report of
  calculation with list of all parameters (technical and physical ones).

  The file {\sf hst.\#}  contains filled histograms.

     The {\it\bf User's menu} serves for an
     implementation of any user's calls.
    For example, it can provide
the interface with the CERN PDF library.

The {\it\bf Users' functions} subroutine allows the user
to implement some
functions written by himself in order

\begin{itemize}
\item  to introduce cuts for any functions of kinematical variables;

\item to make convolution of squared matrix elements with
     any structure functions with account of the  running (strong) coupling
    constant and the momentum transfer  scale;

\item to save  any users' parameters in the file {\sf SESSION.DAT} when
    the program is over and to read them from this file
    when the program starts.
\end{itemize}

\begin{center}
    {\bf 6. Brief review of physical results \\
     obtained by means of the CompHEP package}
\end{center}
\vspace{.5cm}

    Ten three-body processes in the $e^+e^-$ collisions for a heavy particle
       production
such as Higgs, $t$-quark, $W$ and $Z$ are calculated in Ref.\cite{r3}
 by two independent computer codes (generated by CompHEP\cite{r1} and
GRACE\cite{r2}).
The results are in an excellent
agreement
within statistical errors of numerical integration (about 0.5\%). This
 cross-check of numerical results demonstrates that
CompHEP and GRACE systems are quite reliable for a theoretical study of
processes at future
$e^+e^-, e\gamma$ and
$\gamma\gamma$ colliders.

Cross sections of the Higgs boson associated production in $\gamma e$
collisions
are calculated in Ref.\cite{r4} for the $\gamma e\to \nu W H$ and
$\gamma e\to
e Z H$ processes. Event signatures for Higgs boson production, event
separation and
background conditions
are considered. It is shown that the Higgs boson production process $\gamma
e\to
\nu W H$ seems very promising for the investigation of gauge cancellations
between different diagrams and search for anomalous phenomena (for instance,
anomalous Higgs boson interaction vertices).

   In the paper\cite{r5} the calculations of
   total cross sections
    for the $W$
and $Z$ boson production in $\gamma e$ and $\gamma\gamma$ collisions
are presented in the 3rd order
 in
electroweak coupling constant at the tree level. They are compared with the
estimations
obtained
by simple approximation methods to see their accuracy for this class of
processes.
The preliminary physical analysis of obtained results is given.

 All tree level diagrams are  calculated by means of the CompHEP package for
the
 reaction $\gamma\gamma\to t \bar{t} H$ \cite{rg}.
 It was shown that the reaction is very sensitive to probing the Higgs
 fermion coupling in TeV energy range.

   In the paper\cite{r6} the complete tree level calculations for three
particle
    final
state production at  future $e^+e^-, \gamma e$ and $\gamma\gamma$
colliders are
presented.
The results obtained with the help of the CompHEP package for total cross
sections
and
other characteristics of processes in the energy range 0.1-2 TeV are summarized
and their comparison with
the results of other approaches is discussed. In particular
the processes of $W,Z$ and $H$
boson production are considered. These reactions are especially interesting
in connection with
probing new couplings, searching for new particle signals and estimating
the most important backgrounds in
various experiments.

   The possibility of the single and pair excited neutrino production in high
energy $e^+e^-, \gamma e$ and $\gamma\gamma$ collisions at linear colliders
 is studied in Ref.\cite{r7}.
The integrated cross sections of these subprocesses are calculated
in a symbolical form. A special
attention is paid
to a search for excited neutrino in the $\gamma e\to W,W,e$ process. The lower
 limits
for the compositeness parameter to be available in the experiments
at Next Linear Colliders  are
estimated.

    The possibility to detect the Higgs boson signal in the process
$e^+e^-\to Z \bar{b} b$ at LEP200 energies is considered in Ref.\cite{r8}.
 The calculations are performed in the
tree approximation for a complete set of diagrams. Tree level corrections
to the Higgs signal are computed. If the highest possible
LEP200 energy is
$\sqrt{s}$ = 190 GeV the Higgs signal will be very clean for the masses of
Higgs up to $\sqrt{s}-M_Z$ about 95 GeV.

   In \cite{r9} the possibility of Higgs boson signal observation
   at LEP200 and Next Linear Colliders in the reactions
   $e^+e^- \to \mu^+\mu^- b \bar b$, $e^+e^-\to\nu
   \bar{\nu} b \bar{b}$, $e^+e^-\to e^+e^-b \bar b$ is investigated.
   Complete tree level calculations for these $2 \to 4$ processes
   are performed and compared with the various effective $2 \to 2$
   body approximations. The accuracy of effective approximations near
   the thresholds and at different energies are calculated. In some
   situations it is necessary to introduce nontrivial kinematical
   cuts in order to separate the signal from the background.

A complete tree-level calculation of the reaction
$e^+e^-\to e \nu \bar{t} b$
in the electroweak standard theory in the LEP200 energy range is presented
in Ref.\cite{r10}. For top quark masses
in the range 130 to 190 GeV the cross sections are found to be of the order
$10^{-5}-10^{-6}$pb. Therefore, the number of single top quark events is
expected to be negligible
even with an integrated luminosity of L=500 pb$^{-1}$. It is further
demonstrated that the Weizsaecker-Williams
approximation approaches the accurate cross section calculations
reasonably well.

In Ref.\cite{r11} the possibilities of search for vector
leptoquarks at high energy $e p$ and $\gamma p$
colliders are investigated. The exact analytical expressions are derived
for cross sections with the help of CompHEP taking into account
possible anomalous couplings of vector leptoquark with  gauge bosons.
The vector leptoquark search potential at HERA and future $e p$ colliders
is discussed in detail.

Different $2\to3$ reactions for the Higgs production in association
with a vector boson pair at future $e^+e^-$ colliders are calculated
 in the paper
 \cite{rCuypers} using the amplitude technique and the CompHEP  package.
A very good agreement
of two independent calculations has been found. The paper demonstrates
an important point of the CompHEP application for an additional test
of results obtained by other methods or computer systems.

\begin{center}
{\bf 7. Acknowledgement}
\end{center}
The authors express their deep gratitude to
Professor H. S. Song
and to the staff of the Center of Theoretical Physics, Seoul National
University for the hospitality and providing them with excellent conditions
of stay and work in Korea.

\newpage
\framebox{
\begin {tabular}{|c|}\hline
                {\bf CompHEP}\\
                {\bf symbolic module}\\ \hline
                \\
                {\sf Lagrangian (SM and beyond)}\\
                $\downarrow$\\
                {\sf squared matrix element}
                \\
                \\ \hline
\end{tabular}}
\hskip -0.5cm \raise -1.3cm
\vbox{$\Longrightarrow$\framebox{\it Feynman diagrams}
      \vskip 6 mm
      $\Longrightarrow$\framebox{\it Symbolic answer}
      \vskip 6 mm
      $\Longrightarrow$\framebox{
              \begin{tabular}{c}
                       \it Numerical calculator \\
                       (\large \it cross section, distributions)\\
                       \large \it for 1 $\rightarrow$2, 2 $\rightarrow$2
              \end{tabular}
                                }
     }

\vskip 4mm
\hskip 3.5cm $\Downarrow$
\vskip 4mm

\hskip 1cm \framebox{\begin {tabular}{c}
                         \it FORTRAN code\\
                         \it for squared matrix element
                     \end{tabular}
                    }

\vskip 4mm
\hskip 3.5cm $\Downarrow$
\vskip 4mm

\hskip 1.5cm
\framebox{
\begin {tabular}{|c|}\hline
          {\bf CompHEP}\\
          {\bf kinematics module}\\ \hline
          \\
          {\sf Kinematics}\\
          {\sf Cuts}\\
          {\sf Regularizations}
          \\
          \\ \hline
\end{tabular}}
\hskip -0.5cm  \raise -0.5cm
\vbox{$\Longleftarrow$ \framebox{\sf Structure functions}
   \vskip 1cm
   $\Longleftarrow$\framebox{\begin{tabular}{c}
                                 {\bf Monte-Carlo}\\ \hline
                                 BASES (integration)\\
                                 SPRING (event generation)\\
                             \end{tabular}
                            }
     }

\vskip 4mm
\hskip 1.5cm $\Downarrow$ \hskip 3.5cm $\Downarrow$
\vskip 4mm

\framebox{\it Cross section} \hskip 0.5cm
\framebox{\begin{tabular}{c}
                    \it Event flow \\
                    \it Histograms
          \end{tabular}
         }

\vskip 1cm
\centerline{Fig. 1.The  general structure of the CompHEP  package}

\newpage

\small
\begin{picture}(400,500)(0,0)

\put(125,465){
\vbox{
\begin {tabular}{|l|}
 \multicolumn{1}{c}{\rm menu 1}\\ \hline
                 QED  \\
                 Fermi model\\
                 St. model (unit. gauge)\\
                 St. model (Feyn. gauge) \\
                 NEW MODEL \\
                \hline
\end{tabular}

    }
}

\put(200,425){\vector(0,-1){25}}

\put(140,375){
\vbox{
\begin {tabular}{|l|}
\multicolumn{1}{c}{\rm menu2}\\ \hline
                 Enter process   \\
                 Edit model  \\
                 Delete changes \\
                \hline
\end{tabular}

}

}

\put(150,372){\vector(-1,0){80}}
\put(243,382){\vector(1,0){65}}

\put(0,345){
\begin {tabular}{|l|}
 \multicolumn{1}{c}{\rm menu3}\\ \hline
                 Variables \\
                 Constraints\\
                 Particles\\
                 Lagrangian \\
                \hline
\end{tabular}
}

\put(300,367){
\begin {tabular}{|l|}
 \multicolumn{1}{c}{\rm menu 4}\\ \hline
                 Squaring  \\
                 View diagrams\\
                \hline
\end{tabular}
}

\put(297,370){\line(-1,0){10}}
\put(287,370){\vector(0,-1){50}}
\put(235,265){
\begin {tabular}{|l|}
 \multicolumn{1}{c}{\rm menu 5}\\ \hline
                 View squared diagrams  \\
                 Symbolic calculation\\
                 Write results\\
                 REDUCE program \\
                 Numerical calculator\\
                 Enter new process\\
                 Interface \\
                \hline
\end{tabular}
}

\put(235,278){\vector(-1,0){150}}
\put(235,250){\line(-1,0){15}}
\put(220,250){\vector(0,-1){45}}

\put(0,250){
\begin {tabular}{|l|}
 \multicolumn{1}{c}{\rm menu 6}\\ \hline
                 FORTRAN code  \\
                 REDUCE code\\
                 MATHEMATICA code\\
                \hline
\end{tabular}
}

\put(150,165){
\begin {tabular}{|l|}
 \multicolumn{1}{c}{\rm menu 7}\\ \hline
                 View/change data \\
                 (Set angular range) \\
                 (Set precision) \\
                 (Angular dependence) \\
                 Parameter dependence\\
                \hline
\end{tabular}
}

\put(150,136){\vector(-1,0){55}}
\put(275,149){\vector(1,0){60}}

\put(315,129){
\begin {tabular}{|l|}
 \multicolumn{1}{c}{\rm menu 8}\\ \hline
                 Show plot \\
                 Save results in a file\\
                 Recalculate \\
                \hline
\end{tabular}
}

\put(0,120){
\begin {tabular}{|l|}
 \multicolumn{1}{c}{\rm menu 9}\\ \hline
                 (Total cross section) \\
                 (Asymmetry) \\
                \hline
\end{tabular}
}

\put(50,100){\vector(0,-1){30}}
\put(0,50){
\begin {tabular}{|l|}
 \multicolumn{1}{c}{\rm menu 10}\\ \hline
                 Show plot   \\
                 Save results in a file\\
                \hline
\end{tabular}
}

\put(0,0){
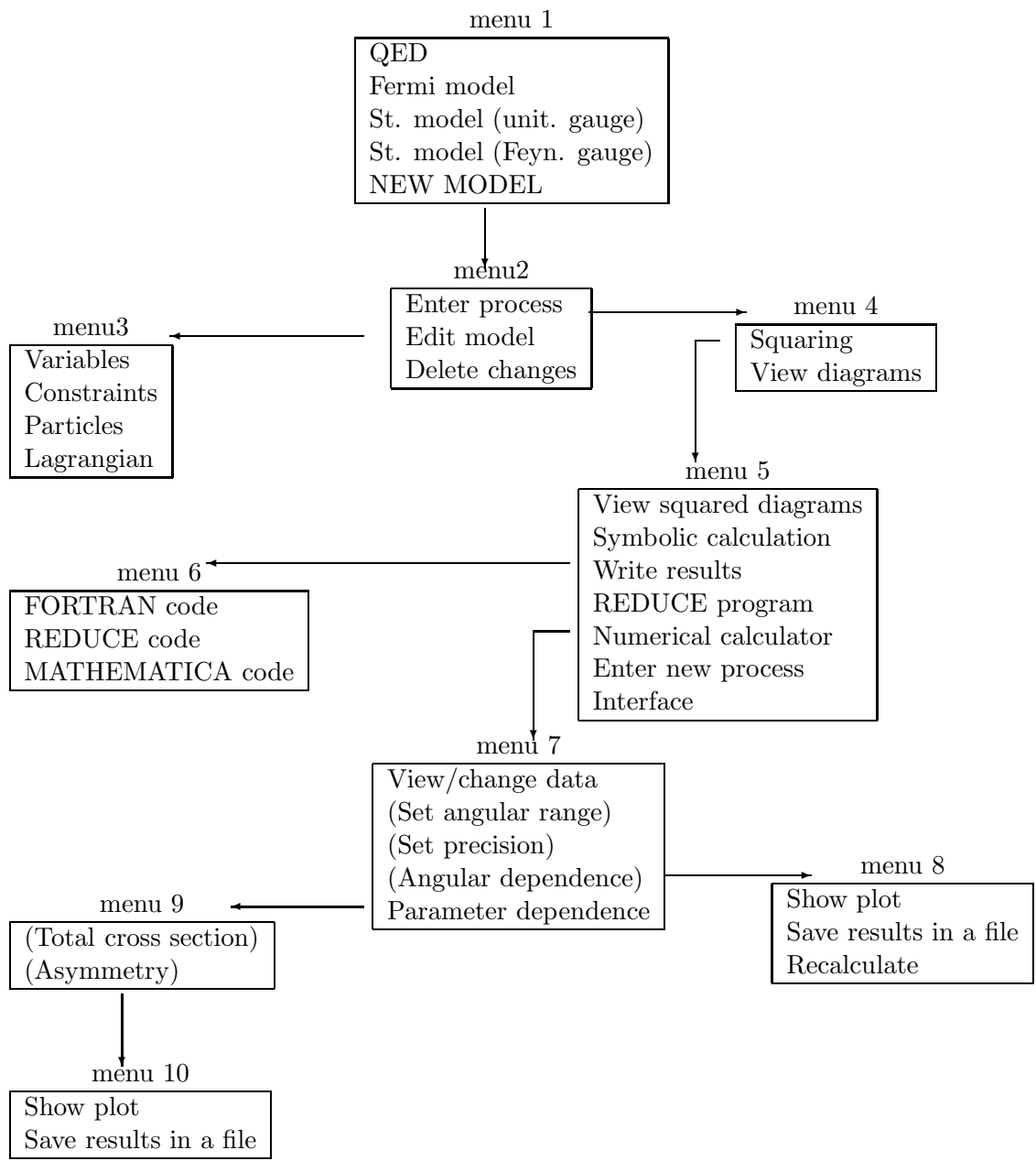
\centerline{Fig. 2. The  menu system for the CompHEP  symbolic part}
}

\end{picture}

\newpage

\def\diagA{
%\begin{figure}
\begin{picture}(170,300)(0,0)
%\label{d1-1}
%\put(85,5){\makebox(5,5)[b]{fig.\ref{d1-1}}}
\put(85,5){\makebox(0,0)[b]{diag 1}}
%\put(85,5){\makebox(0,0)[b]{$e1,u -> e1,u,Z$}}
\thicklines
%\put(0,0){\framebox(170,300){}}
\thinlines
\put(55.0,175.7){\vector(1,0){0}}
\put(34.3,175.7){\makebox(0,0)[r]{$e1$}}
\put(37.1,175.7){\line(1,0){35.7}}
\put(90.7,175.7){\vector(1,0){0}}
\put(90.0,184.3){\makebox(0,0){$e1$}}
\put(72.9,175.7){\line(1,0){35.7}}
\put(147.1,211.4){\makebox(0,0)[l]{$Z$}}
\multiput(108.1,175.7)(4.7,4.7){8}{\rule[-0.5pt]{1.0pt}{1.0pt}}
\multiput(108.6,176.2)(4.7,4.7){8}{\rule[-0.5pt]{1.0pt}{1.0pt}}
\multiput(109.1,176.7)(4.7,4.7){8}{\rule[-0.5pt]{1.0pt}{1.0pt}}
\multiput(109.6,177.2)(4.7,4.7){8}{\rule[-0.5pt]{1.0pt}{1.0pt}}
\multiput(110.1,177.8)(4.7,4.7){8}{\rule[-0.5pt]{1.0pt}{1.0pt}}
\put(126.4,157.9){\vector(1,-1){0}}
\put(147.1,140.0){\makebox(0,0)[l]{$e1$}}
\put(108.6,175.7){\line(1,-1){35.7}}
\put(67.1,140.0){\makebox(0,0)[r]{$A$}}
\multiput(72.4,175.7)(0.0,-4.9){15}{\rule[-0.5pt]{1.0pt}{1.0pt}}
\multiput(72.4,175.2)(0.0,-4.9){15}{\rule[-0.5pt]{1.0pt}{1.0pt}}
\multiput(72.4,174.7)(0.0,-4.9){15}{\rule[-0.5pt]{1.0pt}{1.0pt}}
\multiput(72.4,174.2)(0.0,-4.9){15}{\rule[-0.5pt]{1.0pt}{1.0pt}}
\multiput(72.4,173.7)(0.0,-4.9){15}{\rule[-0.5pt]{1.0pt}{1.0pt}}
\put(55.0,104.3){\vector(1,0){0}}
\put(34.3,104.3){\makebox(0,0)[r]{$u$}}
\put(37.1,104.3){\line(1,0){35.7}}
\put(72.9,104.3){\line(1,0){35.7}}
\put(126.4,86.4){\vector(1,-1){0}}
\put(147.1,68.6){\makebox(0,0)[l]{$u$}}
\put(108.6,104.3){\line(1,-1){35.7}}
\end{picture}
%\end{figure}
}

% CompHEP  version  3.0
% e1,u -> e1,u,Z, diagram # 1-2
\def\diagB{
%\begin{figure}
\begin{picture}(170,300)(0,0)
%\label{d1-2}
%\put(85,5){\makebox(5,5)[b]{fig.\ref{d1-2}}}
\put(85,5){\makebox(0,0)[b]{diag 2}}
%\put(85,5){\makebox(0,0)[b]{$e1,u -> e1,u,Z$}}
\thicklines
%\put(0,0){\framebox(170,300){}}
\thinlines
\put(55.0,175.7){\vector(1,0){0}}
\put(34.3,175.7){\makebox(0,0)[r]{$e1$}}
\put(37.1,175.7){\line(1,0){35.7}}
\put(72.9,175.7){\line(1,0){35.7}}
\put(126.4,193.6){\vector(1,1){0}}
\put(147.1,211.4){\makebox(0,0)[l]{$e1$}}
\put(108.6,175.7){\line(1,1){35.7}}
\put(67.1,140.0){\makebox(0,0)[r]{$A$}}
\multiput(72.4,175.7)(0.0,-4.9){15}{\rule[-0.5pt]{1.0pt}{1.0pt}}
\multiput(72.4,175.2)(0.0,-4.9){15}{\rule[-0.5pt]{1.0pt}{1.0pt}}
\multiput(72.4,174.7)(0.0,-4.9){15}{\rule[-0.5pt]{1.0pt}{1.0pt}}
\multiput(72.4,174.2)(0.0,-4.9){15}{\rule[-0.5pt]{1.0pt}{1.0pt}}
\multiput(72.4,173.7)(0.0,-4.9){15}{\rule[-0.5pt]{1.0pt}{1.0pt}}
\put(55.0,104.3){\vector(1,0){0}}
\put(34.3,104.3){\makebox(0,0)[r]{$u$}}
\put(37.1,104.3){\line(1,0){35.7}}
\put(90.7,104.3){\vector(1,0){0}}
\put(90.0,112.9){\makebox(0,0){$u$}}
\put(72.9,104.3){\line(1,0){35.7}}
\put(147.1,140.0){\makebox(0,0)[l]{$Z$}}
\multiput(108.1,104.3)(4.7,4.7){8}{\rule[-0.5pt]{1.0pt}{1.0pt}}
\multiput(108.6,104.8)(4.7,4.7){8}{\rule[-0.5pt]{1.0pt}{1.0pt}}
\multiput(109.1,105.3)(4.7,4.7){8}{\rule[-0.5pt]{1.0pt}{1.0pt}}
\multiput(109.6,105.8)(4.7,4.7){8}{\rule[-0.5pt]{1.0pt}{1.0pt}}
\multiput(110.1,106.3)(4.7,4.7){8}{\rule[-0.5pt]{1.0pt}{1.0pt}}
\put(126.4,86.4){\vector(1,-1){0}}
\put(147.1,68.6){\makebox(0,0)[l]{$u$}}
\put(108.6,104.3){\line(1,-1){35.7}}
\end{picture}
%\end{figure}
}
\def\diagC{
% CompHEP  version  3.0
% e1,u -> e1,u,Z, diagram # 1-3
%\begin{figure}
\begin{picture}(170,300)(0,0)
%\label{d1-3}
%\put(85,5){\makebox(5,5)[b]{fig.\ref{d1-3}}}
\put(85,5){\makebox(0,0)[b]{diag 3}}
%\put(85,5){\makebox(0,0)[b]{$e1,u -> e1,u,Z$}}
\thicklines
%\put(0,0){\framebox(170,300){}}
\thinlines
\put(72.9,211.4){\vector(1,0){0}}
\put(34.3,211.4){\makebox(0,0)[r]{$e1$}}
\put(37.1,211.4){\line(1,0){71.4}}
\put(126.4,211.4){\vector(1,0){0}}
\put(147.1,211.4){\makebox(0,0)[l]{$e1$}}
\put(108.6,211.4){\line(1,0){35.7}}
\put(102.9,175.7){\makebox(0,0)[r]{$A$}}
\multiput(108.1,211.4)(0.0,-4.9){15}{\rule[-0.5pt]{1.0pt}{1.0pt}}
\multiput(108.1,210.9)(0.0,-4.9){15}{\rule[-0.5pt]{1.0pt}{1.0pt}}
\multiput(108.1,210.4)(0.0,-4.9){15}{\rule[-0.5pt]{1.0pt}{1.0pt}}
\multiput(108.1,209.9)(0.0,-4.9){15}{\rule[-0.5pt]{1.0pt}{1.0pt}}
\multiput(108.1,209.4)(0.0,-4.9){15}{\rule[-0.5pt]{1.0pt}{1.0pt}}
\put(126.4,140.0){\vector(1,0){0}}
\put(147.1,140.0){\makebox(0,0)[l]{$u$}}
\put(108.6,140.0){\line(1,0){35.7}}
\put(108.6,104.3){\vector(0,1){0}}
\put(102.9,104.3){\makebox(0,0)[r]{$u$}}
\put(108.6,140.0){\line(0,-1){71.4}}
\put(72.9,68.6){\vector(1,0){0}}
\put(34.3,68.6){\makebox(0,0)[r]{$u$}}
\put(37.1,68.6){\line(1,0){71.4}}
\put(147.1,68.6){\makebox(0,0)[l]{$Z$}}
\multiput(108.1,68.6)(4.7,0.0){8}{\rule[-0.5pt]{1.0pt}{1.0pt}}
\multiput(108.6,68.6)(4.7,0.0){8}{\rule[-0.5pt]{1.0pt}{1.0pt}}
\multiput(109.1,68.6)(4.7,0.0){8}{\rule[-0.5pt]{1.0pt}{1.0pt}}
\multiput(109.6,68.6)(4.7,0.0){8}{\rule[-0.5pt]{1.0pt}{1.0pt}}
\multiput(110.1,68.6)(4.7,0.0){8}{\rule[-0.5pt]{1.0pt}{1.0pt}}
\end{picture}
%\end{figure}
}
\def\diagD{
% CompHEP  version  3.0
% e1,u -> e1,u,Z, diagram # 1-4
%\begin{figure}
\begin{picture}(170,300)(0,0)
%\label{d1-4}
%\put(85,5){\makebox(5,5)[b]{fig.\ref{d1-4}}}
\put(85,5){\makebox(0,0)[b]{diag 4}}
%\put(85,5){\makebox(0,0)[b]{$e1,u -> e1,u,Z$}}
\thicklines
%\put(0,0){\framebox(170,300){}}
\thinlines
\put(55.0,175.7){\vector(1,0){0}}
\put(34.3,175.7){\makebox(0,0)[r]{$e1$}}
\put(37.1,175.7){\line(1,0){35.7}}
\put(72.9,175.7){\line(1,0){35.7}}
\put(126.4,193.6){\vector(1,1){0}}
\put(147.1,211.4){\makebox(0,0)[l]{$e1$}}
\put(108.6,175.7){\line(1,1){35.7}}
\put(67.1,140.0){\makebox(0,0)[r]{$Z$}}
\multiput(72.4,175.7)(0.0,-4.9){15}{\rule[-0.5pt]{1.0pt}{1.0pt}}
\multiput(72.4,175.2)(0.0,-4.9){15}{\rule[-0.5pt]{1.0pt}{1.0pt}}
\multiput(72.4,174.7)(0.0,-4.9){15}{\rule[-0.5pt]{1.0pt}{1.0pt}}
\multiput(72.4,174.2)(0.0,-4.9){15}{\rule[-0.5pt]{1.0pt}{1.0pt}}
\multiput(72.4,173.7)(0.0,-4.9){15}{\rule[-0.5pt]{1.0pt}{1.0pt}}
\put(55.0,104.3){\vector(1,0){0}}
\put(34.3,104.3){\makebox(0,0)[r]{$u$}}
\put(37.1,104.3){\line(1,0){35.7}}
\put(90.7,104.3){\vector(1,0){0}}
\put(90.0,112.9){\makebox(0,0){$u$}}
\put(72.9,104.3){\line(1,0){35.7}}
\put(147.1,140.0){\makebox(0,0)[l]{$Z$}}
\multiput(108.1,104.3)(4.7,4.7){8}{\rule[-0.5pt]{1.0pt}{1.0pt}}
\multiput(108.6,104.8)(4.7,4.7){8}{\rule[-0.5pt]{1.0pt}{1.0pt}}
\multiput(109.1,105.3)(4.7,4.7){8}{\rule[-0.5pt]{1.0pt}{1.0pt}}
\multiput(109.6,105.8)(4.7,4.7){8}{\rule[-0.5pt]{1.0pt}{1.0pt}}
\multiput(110.1,106.3)(4.7,4.7){8}{\rule[-0.5pt]{1.0pt}{1.0pt}}
\put(126.4,86.4){\vector(1,-1){0}}
\put(147.1,68.6){\makebox(0,0)[l]{$u$}}
\put(108.6,104.3){\line(1,-1){35.7}}
\end{picture}
%\end{figure}
}
\def\diagE{
% CompHEP  version  3.0
% e1,u -> e1,u,Z, diagram # 1-5
%\begin{figure}
\begin{picture}(170,300)(0,0)
%\label{d1-5}
%\put(85,5){\makebox(5,5)[b]{fig.\ref{d1-5}}}
\put(85,5){\makebox(0,0)[b]{diag 5}}
%\put(85,5){\makebox(0,0)[b]{$e1,u -> e1,u,Z$}}
\thicklines
%\put(0,0){\framebox(170,300){}}
\thinlines
\put(72.9,211.4){\vector(1,0){0}}
\put(34.3,211.4){\makebox(0,0)[r]{$e1$}}
\put(37.1,211.4){\line(1,0){71.4}}
\put(126.4,211.4){\vector(1,0){0}}
\put(147.1,211.4){\makebox(0,0)[l]{$e1$}}
\put(108.6,211.4){\line(1,0){35.7}}
\put(102.9,175.7){\makebox(0,0)[r]{$Z$}}
\multiput(108.1,211.4)(0.0,-4.9){15}{\rule[-0.5pt]{1.0pt}{1.0pt}}
\multiput(108.1,210.9)(0.0,-4.9){15}{\rule[-0.5pt]{1.0pt}{1.0pt}}
\multiput(108.1,210.4)(0.0,-4.9){15}{\rule[-0.5pt]{1.0pt}{1.0pt}}
\multiput(108.1,209.9)(0.0,-4.9){15}{\rule[-0.5pt]{1.0pt}{1.0pt}}
\multiput(108.1,209.4)(0.0,-4.9){15}{\rule[-0.5pt]{1.0pt}{1.0pt}}
\put(126.4,140.0){\vector(1,0){0}}
\put(147.1,140.0){\makebox(0,0)[l]{$u$}}
\put(108.6,140.0){\line(1,0){35.7}}
\put(108.6,104.3){\vector(0,1){0}}
\put(102.9,104.3){\makebox(0,0)[r]{$u$}}
\put(108.6,140.0){\line(0,-1){71.4}}
\put(72.9,68.6){\vector(1,0){0}}
\put(34.3,68.6){\makebox(0,0)[r]{$u$}}
\put(37.1,68.6){\line(1,0){71.4}}
\put(147.1,68.6){\makebox(0,0)[l]{$Z$}}
\multiput(108.1,68.6)(4.7,0.0){8}{\rule[-0.5pt]{1.0pt}{1.0pt}}
\multiput(108.6,68.6)(4.7,0.0){8}{\rule[-0.5pt]{1.0pt}{1.0pt}}
\multiput(109.1,68.6)(4.7,0.0){8}{\rule[-0.5pt]{1.0pt}{1.0pt}}
\multiput(109.6,68.6)(4.7,0.0){8}{\rule[-0.5pt]{1.0pt}{1.0pt}}
\multiput(110.1,68.6)(4.7,0.0){8}{\rule[-0.5pt]{1.0pt}{1.0pt}}
\end{picture}
%\end{figure}
}
\def\diagF{
% CompHEP  version  3.0
% e1,u -> e1,u,Z, diagram # 1-6
%\begin{figure}
\begin{picture}(170,300)(0,0)
%\label{d1-6}
%\put(85,5){\makebox(5,5)[b]{fig.\ref{d1-6}}}
\put(85,5){\makebox(0,0)[b]{diag 6}}
%\put(85,5){\makebox(0,0)[b]{$e1,u -> e1,u,Z$}}
\thicklines
%\put(0,0){\framebox(170,300){}}
\thinlines
\put(55.0,175.7){\vector(1,0){0}}
\put(34.3,175.7){\makebox(0,0)[r]{$e1$}}
\put(37.1,175.7){\line(1,0){35.7}}
\put(90.7,175.7){\vector(1,0){0}}
\put(90.0,184.3){\makebox(0,0){$e1$}}
\put(72.9,175.7){\line(1,0){35.7}}
\put(147.1,211.4){\makebox(0,0)[l]{$Z$}}
\multiput(108.1,175.7)(4.7,4.7){8}{\rule[-0.5pt]{1.0pt}{1.0pt}}
\multiput(108.6,176.2)(4.7,4.7){8}{\rule[-0.5pt]{1.0pt}{1.0pt}}
\multiput(109.1,176.7)(4.7,4.7){8}{\rule[-0.5pt]{1.0pt}{1.0pt}}
\multiput(109.6,177.2)(4.7,4.7){8}{\rule[-0.5pt]{1.0pt}{1.0pt}}
\multiput(110.1,177.8)(4.7,4.7){8}{\rule[-0.5pt]{1.0pt}{1.0pt}}
\put(126.4,157.9){\vector(1,-1){0}}
\put(147.1,140.0){\makebox(0,0)[l]{$e1$}}
\put(108.6,175.7){\line(1,-1){35.7}}
\put(67.1,140.0){\makebox(0,0)[r]{$Z$}}
\multiput(72.4,175.7)(0.0,-4.9){15}{\rule[-0.5pt]{1.0pt}{1.0pt}}
\multiput(72.4,175.2)(0.0,-4.9){15}{\rule[-0.5pt]{1.0pt}{1.0pt}}
\multiput(72.4,174.7)(0.0,-4.9){15}{\rule[-0.5pt]{1.0pt}{1.0pt}}
\multiput(72.4,174.2)(0.0,-4.9){15}{\rule[-0.5pt]{1.0pt}{1.0pt}}
\multiput(72.4,173.7)(0.0,-4.9){15}{\rule[-0.5pt]{1.0pt}{1.0pt}}
\put(55.0,104.3){\vector(1,0){0}}
\put(34.3,104.3){\makebox(0,0)[r]{$u$}}
\put(37.1,104.3){\line(1,0){35.7}}
\put(72.9,104.3){\line(1,0){35.7}}
\put(126.4,86.4){\vector(1,-1){0}}
\put(147.1,68.6){\makebox(0,0)[l]{$u$}}
\put(108.6,104.3){\line(1,-1){35.7}}
\end{picture}
%\end{figure}
}
\def\diagG{
% CompHEP  version  3.0
% e1,u -> e1,u,Z, diagram # 1-7
%\begin{figure}
\begin{picture}(170,300)(0,0)
%\label{d1-7}
%\put(85,5){\makebox(5,5)[b]{fig.\ref{d1-7}}}
\put(85,5){\makebox(0,0)[b]{diag 7}}
%\put(85,5){\makebox(0,0)[b]{$e1,u -> e1,u,Z$}}
\thicklines
%\put(0,0){\framebox(170,300){}}
\thinlines
\put(72.9,211.4){\vector(1,0){0}}
\put(34.3,211.4){\makebox(0,0)[r]{$e1$}}
\put(37.1,211.4){\line(1,0){71.4}}
\put(147.1,211.4){\makebox(0,0)[l]{$Z$}}
\multiput(108.1,211.4)(4.7,0.0){8}{\rule[-0.5pt]{1.0pt}{1.0pt}}
\multiput(108.6,211.4)(4.7,0.0){8}{\rule[-0.5pt]{1.0pt}{1.0pt}}
\multiput(109.1,211.4)(4.7,0.0){8}{\rule[-0.5pt]{1.0pt}{1.0pt}}
\multiput(109.6,211.4)(4.7,0.0){8}{\rule[-0.5pt]{1.0pt}{1.0pt}}
\multiput(110.1,211.4)(4.7,0.0){8}{\rule[-0.5pt]{1.0pt}{1.0pt}}
\put(108.6,175.7){\vector(0,-1){0}}
\put(102.9,175.7){\makebox(0,0)[r]{$e1$}}
\put(108.6,211.4){\line(0,-1){71.4}}
\put(126.4,140.0){\vector(1,0){0}}
\put(147.1,140.0){\makebox(0,0)[l]{$e1$}}
\put(108.6,140.0){\line(1,0){35.7}}
\put(102.9,104.3){\makebox(0,0)[r]{$A$}}
\multiput(108.1,140.0)(0.0,-4.9){15}{\rule[-0.5pt]{1.0pt}{1.0pt}}
\multiput(108.1,139.5)(0.0,-4.9){15}{\rule[-0.5pt]{1.0pt}{1.0pt}}
\multiput(108.1,139.0)(0.0,-4.9){15}{\rule[-0.5pt]{1.0pt}{1.0pt}}
\multiput(108.1,138.5)(0.0,-4.9){15}{\rule[-0.5pt]{1.0pt}{1.0pt}}
\multiput(108.1,138.0)(0.0,-4.9){15}{\rule[-0.5pt]{1.0pt}{1.0pt}}
\put(72.9,68.6){\vector(1,0){0}}
\put(34.3,68.6){\makebox(0,0)[r]{$u$}}
\put(37.1,68.6){\line(1,0){71.4}}
\put(126.4,68.6){\vector(1,0){0}}
\put(147.1,68.6){\makebox(0,0)[l]{$u$}}
\put(108.6,68.6){\line(1,0){35.7}}
\end{picture}
%\end{figure}
}
\def\diagH{
% CompHEP  version  3.0
% e1,u -> e1,u,Z, diagram # 1-8
%\begin{figure}
\begin{picture}(170,300)(0,0)
%\label{d1-8}
%\put(85,5){\makebox(5,5)[b]{fig.\ref{d1-8}}}
\put(85,5){\makebox(0,0)[b]{diag 8}}
%\put(85,5){\makebox(0,0)[b]{$e1,u -> e1,u,Z$}}
\thicklines
%\put(0,0){\framebox(170,300){}}
\thinlines
\put(72.9,211.4){\vector(1,0){0}}
\put(34.3,211.4){\makebox(0,0)[r]{$e1$}}
\put(37.1,211.4){\line(1,0){71.4}}
\put(147.1,211.4){\makebox(0,0)[l]{$Z$}}
\multiput(108.1,211.4)(4.7,0.0){8}{\rule[-0.5pt]{1.0pt}{1.0pt}}
\multiput(108.6,211.4)(4.7,0.0){8}{\rule[-0.5pt]{1.0pt}{1.0pt}}
\multiput(109.1,211.4)(4.7,0.0){8}{\rule[-0.5pt]{1.0pt}{1.0pt}}
\multiput(109.6,211.4)(4.7,0.0){8}{\rule[-0.5pt]{1.0pt}{1.0pt}}
\multiput(110.1,211.4)(4.7,0.0){8}{\rule[-0.5pt]{1.0pt}{1.0pt}}
\put(108.6,175.7){\vector(0,-1){0}}
\put(102.9,175.7){\makebox(0,0)[r]{$e1$}}
\put(108.6,211.4){\line(0,-1){71.4}}
\put(126.4,140.0){\vector(1,0){0}}
\put(147.1,140.0){\makebox(0,0)[l]{$e1$}}
\put(108.6,140.0){\line(1,0){35.7}}
\put(102.9,104.3){\makebox(0,0)[r]{$Z$}}
\multiput(108.1,140.0)(0.0,-4.9){15}{\rule[-0.5pt]{1.0pt}{1.0pt}}
\multiput(108.1,139.5)(0.0,-4.9){15}{\rule[-0.5pt]{1.0pt}{1.0pt}}
\multiput(108.1,139.0)(0.0,-4.9){15}{\rule[-0.5pt]{1.0pt}{1.0pt}}
\multiput(108.1,138.5)(0.0,-4.9){15}{\rule[-0.5pt]{1.0pt}{1.0pt}}
\multiput(108.1,138.0)(0.0,-4.9){15}{\rule[-0.5pt]{1.0pt}{1.0pt}}
\put(72.9,68.6){\vector(1,0){0}}
\put(34.3,68.6){\makebox(0,0)[r]{$u$}}
\put(37.1,68.6){\line(1,0){71.4}}
\put(126.4,68.6){\vector(1,0){0}}
\put(147.1,68.6){\makebox(0,0)[l]{$u$}}
\put(108.6,68.6){\line(1,0){35.7}}
\end{picture}
%\end{figure}
}
{
\unitlength=0.5pt
\scriptsize
\diagA\ \diagB\ \diagC\ \diagD\

\diagE\ \diagF\ \diagG\ \diagH
\normalsize
\begin{center}
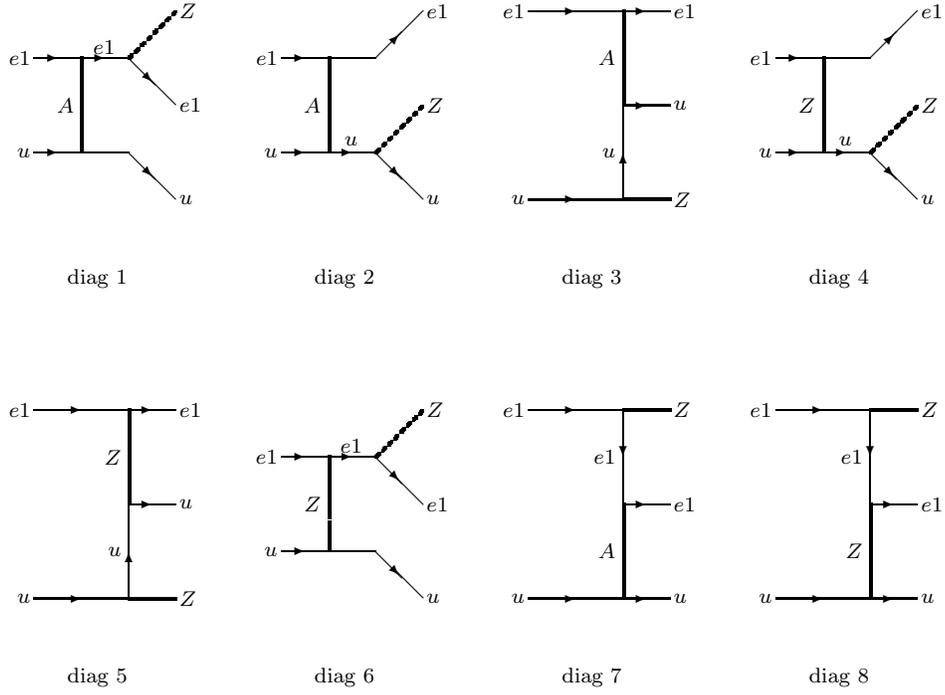
{ Fig. 3. Diagrams for process $ e^-,u -> e^-,u,Z $}
\end{center}
}

\newpage

\bf

\vbox{
\centerline{
\begin {tabular}{|ll|}
 \multicolumn{2}{c}{\rm Main menu }\\ \hline
1. Calculation      &  2. IN state \\
3. Model parameters &  4. Invariant cuts \\
5. Kinematics       &  6. MC parameters \\
7. Regularization   &   8. Task formation \\
9. View results     & 10. User`s menu \\
 \hline
\end{tabular}
}
  }

\centerline{
\begin{picture}(160,100)
\put(0,20){
\vbox{
\begin {tabular}{|l|}
 \multicolumn{1}{c}{\rm In state }\\ \hline
1. StructF(1) = {\it OFF } \\    2. SQRTS = {\it 1000 }   \\
3. StructF(2) = {\it OFF }                 \\
 \hline
\end{tabular}
   }
   }
\end{picture}
\begin{picture}(160,100)
\put(0,20){
\vbox{
\begin {tabular}{|l|}
\multicolumn{1}{c}{\rm Invariant cuts}\\ \hline
1. Insert new cut \\ 2. Delete cut \\
3. Change cut        \\
 \hline
\end{tabular}
  }
  }
\end{picture}
}

\vskip 1.0 cm
\vbox{
\centerline{
\begin {tabular}{|ll|}
 \multicolumn{2}{c}{\rm MC parameters }\\ \hline
1. Ncall = {\it 10000} &  2. Acc1= {\it 0.1 }\\
3. Itmx1= {\it 5 }   & 4. Acc2= {\it 0.1 } \\
5. Itmx2={\it 0}   & 6. Event generator {\rm OFF} \\
7. Number of events = {\it 1000 } &  \\
 \hline
\end{tabular}
  }
}

\centerline{
\begin{picture}(180,100)
\put(0,20){
\vbox{
\begin {tabular}{|l|}
 \multicolumn{1}{c}{\rm Regularization }\\ \hline
1. Insert new regularization \\  2. Delete regularization \\
3. Change regularization       \\
 \hline
\end{tabular}
}
}
\end{picture}
\begin{picture}(160,100)
\put(0,20){
\vbox{
\begin {tabular}{|l|}
 \multicolumn{1}{c}{\rm Task formation }\\ \hline
1. Table parameters  \\  2. Set default session \\
3. Add session to batch  \\
 \hline
\end{tabular}
}
}
\end{picture}
}
\vskip 1.0 cm
\vbox{
\centerline{
\begin {tabular}{|ll|}
 \multicolumn{2}{c}{\rm View results}\\ \hline
1.  session \# to view - {\it 3 } &  2. View result file \\
3. View protocol file      &  4. View histogram file \\
 \hline
\end{tabular}
  }
}
\vskip 1.0 cm
{
\centerline{\rm Fig.4  The menu system for the  CompHEP numerical part}
}


\vspace{1cm}

\begin{thebibliography}{99}

\bibitem{r1} E. Boos, M. Dubinin, V. Edneral, V. Ilyin, A. Kryukov, A.Pukhov,
    V. Savrin, S. Shichanin and A. Taranov.
    in: Proc. Int. Workshop on Software Engineering, Artificial Intelligence
    and Expert Systems for High Energy and Nuclear Physics (AI'90), ed.
    by D. Perret-Gallix and W. Wojcik, Editions du CNRS, 1990, p. 573.

    L. Gladilin, V. Ilyin, A. Pukhov.
    in: Proc. CHEP92, ed.by C.Verkerk and W.Wojcik, CERN 92-07, 1992, p. 855.

    E. Boos, M. Dubinin, V. Edneral, V. Ilyin, A. Kryukov, A. Pukhov,
    S. Shichanin.
    in: "New Computing Techniques in Physics Research II",
    ed.by D. Perret-Gallix, World Scientific, Singapore, 1992, p. 665.

    A. Pukhov.
    in: "New Computing Techniques in Physics Research III",
    ed. by K.-H.Becks and D.Perret-Gallix, World Scientific,
    Singapore, 1993, p. 473.

\bibitem{r2} S. Kawabata.
    Comp. Phys. Comm. 41 (1986) 127.

    S. Kawabata, T. Kaneko.
    Comp. Phys. Comm. 48 (1988) 353.

    T. Ishikawa, T. Kaneko,K. Kato, S. Kawabata, Y. Shimizu and
    H. Tanaka.
    GRACE manual, KEK report 92-19, 1993.


\bibitem{r3} E. Boos, M. Dubinin, V. Ilyin, A. Pukhov, S. Shichanin,
    Y. Shimizu, T. Kaneko, S. Kawabata, Y. Kurihara.
    Int. J. Mod. Phys. C5 (1994) 615-628.


\bibitem{r4} E. Boos, M. Dubinin, V. Ilyin, A. Pukhov, G. Jikia, S.
Sultanov. Phys.Lett. B273 (1991) 173.



\bibitem{r5} I. Ginzburg, V. Ilyin, A. Pukhov, V. Serbo, S. Shichanin.
   Phys. of Atomic Nuclei, 56 (1993) 1481 (Yad.Fiz., 56 (1993) 57)

\bibitem{rg} E. Boos, I. Ginzburg, K. Melnikov, T. Sack, S. Shichanin.
    Z.Phys. C56 (1992) 487.


\bibitem{r6} E. Boos, M. Dubinin, V. Ilyin, A. Pukhov.
   in: $e^+e^-$ collisions at 500 GeV: The physics potential. Proc. of the
    Workshop
   Munich-Annecy-Hamburg, ed. by P.M.Zerwas,
   DESY report 93-123C, 1993, p. 561.


\bibitem{r7}A. Belyaev, E. Boos, A. Pukhov.
   Phys.Lett., B296 (1992) 452.


   A. Belyaev, E. Boos.
  Phys. of Atomic Nuclei, 56 (1993) 1447 (Yad.Fiz., 56 (1993) 5)


\bibitem{r8} E. Boos, M. Dubinin.
   Phys. Lett. B308 (1993) 147.

             E.Boos, M.Dubinin.
   Phys. of Atomic Nuclei, 56 (1993) 1455 (Yad.Fiz., 56 (1993) 16)


\bibitem{r9} E. Boos, M. Sachwitz, H. J. Schreiber, S. Shichanin.
   Z.Phys. C61 (1994) 675.

             E. Boos, M. Sachwitz, H. J. Schreiber, S. Shichanin.
   DESY preprint 93-183 (1993), Int. J. Mod. Phys A (to be published)

             M. Dubinin, V. Edneral, Y. Kurihara, Y. Shimizu.
   Phys.Lett., B329 (1994) 379

             E. Boos, M. Sachwitz, H. J. Schreiber, S. Shichanin.
   DESY preprint 94-091 (1994), Z. Phys. C (to be published)

\bibitem{r10} E. Boos,  V. Ilyin, A. Pukhov, S. Shichanin, M. Sachwitz,
   H. J. Schreiber,
   T. Ishikawa, T. Kaneko, S. Kawabata, Y. Kurihara and Y. Shimizu.
   Phys.Lett., B326 (1994) 190.

\bibitem{r11} J. Blumlein, E. Boos, A. Pukhov.
   DESY preprint 94-072 (1994), Mod. Phys. Lett. (to be published)

\bibitem{rCuypers} M. Baillargeon, F. Boudjema, F. Cuypers, E. Gabrielli,
                   B. Mele.
                   Nucl.Phys. B424 (1994) 343

\end{thebibliography}
\end{document}